\documentclass[aps,showkeys,nofootinbib,reprint,pra,10pt,onecolumn]{revtex4-1}

\usepackage{amsthm,amssymb}
\usepackage{latexsym}
\usepackage{graphicx}
\usepackage{graphics}
\usepackage[T1]{fontenc}
\usepackage{multirow}
\usepackage{enumerate}

\newcommand{\ket}[1]{\mbox{$ | #1 \rangle $}}

\begin{document}

\title{Polarization entangled photon-pair source based on quantum nonlinear photonics and interferometry}

\author{F. Kaiser$^{1,\dagger}$}
\author{L. A. Ngah$^1$}
\author{A. Issautier$^1$}
\author{T. Delord$^{1,2}$}
\author{D. Aktas$^1$}
\author{V. D'Auria$^1$}
\author{M. P. De Micheli$^1$}
\author{A. Kastberg$^1$}
\author{L. Labont\'{e}$^1$}
\author{O. Alibart$^1$}
\author{A. Martin$^{1,\ddagger}$}
\author{S. Tanzilli$^1$}\email{sebastien.tanzilli@unice.fr}
\address{$^1$ Laboratoire de Physique de la Mati\`{e}re Condens\'{e}e, Universit\'{e} Nice Sophia Antipolis,\\CNRS UMR 7336, Parc Valrose, F-06108 Nice Cedex 2, France\\
$^2$ \'Ecole Normale Sup\'erieure de Lyon, 46 All\'ee d'Italie, F-69364 Lyon Cedex 07, France\\
$^\dagger$ Currently with the Deutsches Elektronen-Synchrotron DESY, Hamburg, Germany\\
$^\ddagger$ Currently with the Group of Applied Physics, University of Geneva, Switzerland}

\begin{abstract}
We present a versatile, high-brightness, guided-wave source of polarization entangled photons, emitted at a telecom wavelength. Photon-pairs are generated using an integrated type-0 nonlinear waveguide, and subsequently prepared in a polarization entangled state via a stabilized fiber interferometer. We show that the single photon emission wavelength can be tuned over more than 50\,nm, whereas the single photon spectral bandwidth can be chosen at will over more than five orders of magnitude (from 25\,MHz to 4\,THz). Moreover, by performing entanglement analysis, we demonstrate a high degree of control of the quantum state via the violation of the Bell inequalities by more than 40 standard deviations. This makes this scheme suitable for a wide range of quantum optics experiments, ranging from fundamental research to quantum information applications.\\
We report on details of the setup, as well as on the characterization of all included components, previously outlined in F. Kaiser \textit{et al.} (2013 Laser Phys. Lett. \textbf{10}, 045202).
\end{abstract}


\maketitle

\tableofcontents

\section{Introduction}

Quantum communication science has become a very broad and active field of research. On one hand, quantum key distribution (QKD)~\cite{scarani_2009}, allowing secure distribution of cryptography ciphers between distant partners, has reached the commercial market as well as high-speed system capabablities~\cite{Yoshino_HSQKD_2012}. Related networking protocols, such as entanglement based quantum relay operations, are employed as a means for extending the distance of quantum communication links~\cite{aboussouan_QR_2010,McMillan_2photSeparated_2013}. On the other hand, the study of light/matter interaction is a promising approach for implementing quantum storage devices~\cite{lvovsky_OQM_2009}. Those devices are essential elements to achieve quantum repeater scenarios in which entanglement is distributed, stored, and distilled, all in a heralded fashion, making it possible to increase the overall link efficiency~\cite{Sangouard_Rep_2011}. 
 
Over the past three decades, entanglement has been widely exploited as a resource in fundamental tests of quantum physics~\cite{aspect_experimental_1982,Tittel_2001}. We find, among others, nonlocality tests involving spacelike separated paired photons~\cite{Tittel_Bell10_1998,Weihs_Bell_1998}, quantum delayed-choice experiments~\cite{Ma_2012,peruzzo_2012,kaiser_2012}, and demonstrations of micro-macro entangled states of light~\cite{Lvovsky_MicroMacro_2013,Bruno_MicroMacro_2013}. 
Moreover, with the emergence of long-distance quantum communication links working at telecom wavelengths~\cite{lloyd_long_2001,landry_quantum_2007,Ursin_entanglement_2007,Hubel_2007}, new generation sources have been developed, featuring higher brightness, better stability, compactness, and near perfect entanglement fidelities. Photons can now be generated over narrow enough bandwidths ($\leq$100\,GHz) to avoid both chromatic and polarization mode dispersion along the distribution fibers~\cite{Fasel2004a}. More importantly, current light/matter interaction based quantum memories have, depending on both the physical system and the applied storage protocol, acceptance bandwidths ranging from some MHz to several GHz~\cite{lvovsky_OQM_2009}.
Consequently, to push long-distance quantum communication one step further, there is a need for implementing versatile solutions so as to benefit from the advantages of different quantum technologies. In this framework, sources based on quantum integrated photonics~\cite{Tanzilli_Genesis_2012} appear to be natural and very promising candidates, offering the possibility to efficiently create polarization entanglement at wavelengths compatible with standard fiber components~\cite{kaiserI_2012a,Herrmann2013a}.

In the following, we present the details of a versatile, high-brightness, source of polarization entangled photons, whose main results were first presented in Ref.~\cite{Kaiser_type0_2013}. Its key features, \textit{i.e.}, the central emission wavelength, the spectral bandwidth, and the quantum state, can be tuned at will and adapted to a broad range of quantum network applications. This is enabled by a pertinent combination of an integrated nonlinear optics photon-pair generator, standard telecommunication components, and an entangled state preparation stage based on a stabilized Mach-Zehnder interferometer (MZI). After the presentation of the overall setup and the principle of the source, we will detail all the key elements, namely the integrated nonlinear generator, employed filters, and the quantum state preparation stage. We will then present the entanglement characterization proving the relevance of our approach. Related stabilization schemes and performances in terms of brightness and internal losses will also be outlined. Finally, we will summarize the obtained results, in the perspective of overall performances, and discuss potential improvements.

\section{Specifications, setup, and principle of the source}

Generation of polarization entanglement has been demonstrated using various strategies based on nonlinear media, such as microstructured fibers~\cite{Fulconis_PolarFiber_2007,Fang_PolarFiber_2013}, single pass bulk crystals~\cite{Fiorentino2008,piro_2009}, crystals surrounded by a cavity referred to as optical parametric oscillators (OPO) below the threshold~\cite{Wang2004,Kuklewicz_2006,bao_2008}, or type-II waveguide crystals~\cite{kaiserI_2012a,Herrmann2013a}. However, these strategies have all shown relatively low brightness, which becomes an issue when (ultra-)narrowband photons are needed. Other approaches, capable of generating narrowband polarization entangled photons based on quantum dots~\cite{dousse_2010} or cold atomic ensembles~\cite{dudin_2010,yan_2011} have recently been demonstrated, albeit showing limited entanglement fidelities. 

Our source specifications are outlined in the following.
First, the paired photons are emitted at a wavelength lying in the telecom C-band (1530 - 1565\,nm), and further collected using a single mode telecom fiber in order to benefit from both standard components for routing and filtering purposes and low propagation losses in case of distribution over a long distance.
Second, the photon bandwidth is made readily adaptable so as to be compatible with a broad variety of applications, ranging from QKD in telecommunication channels to quantum storage device implementations.
Third, the coding of quantum information relies on the polarization observable since entanglement correlations can be measured using simple analyzers, being free of interferometric devices as opposed to the case of the time-bin observable~\cite{Martin_TB_2013}. In addition, polarization entanglement can now be distributed over long distances thanks to active compensation systems of fiber birefringence fluctuations~\cite{xavier_2009}.
Eventually, and importantly, the key figures of merit are a high rate of available photon-pairs, and a fidelity to the desired entangled state as close to unity as possible.

\begin{figure}[h!]
\center
\includegraphics[width=0.75\columnwidth]{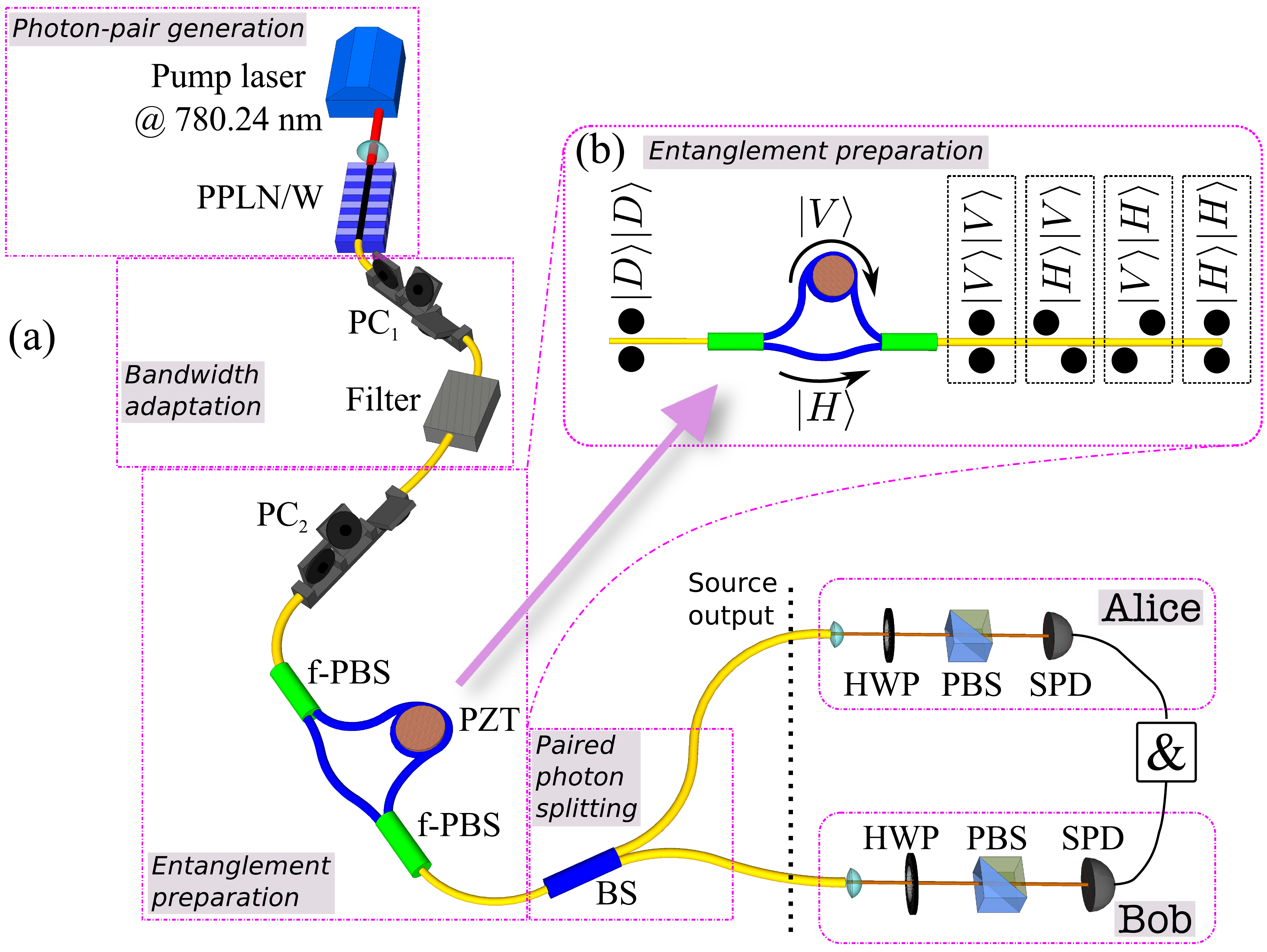}
\caption{\label{Fig_Exp}Experimental setup. (a) Light from a stabilized laser at 780.24\,nm is sent through a PPLN/W for the generation of twin photons at the degenerate wavelength 1560.48\,nm. After collection with a single mode telecom fiber, the paired photons pass through a polarization controller (PC$_1$) followed by a filter, adapting their bandwidth to the desired application. With commercially available components, the filtering transmission bandwidth can be chosen from some MHz to several hundreds of GHz. An unbalanced MZI made, of two fiber-polarizing beam-splitters (f-PBS) connected by polarization maintaining fibers, allows preparation of the desired polarization entangled state (via post-selection in the time domain). A fiber beam-splitter (BS) is used to separate the photons and to distribute them to Alice and Bob. At both locations, measurements are carried out using a polarization state analyzer, comprising a half-wave plate (HWP), a polarizing beam-splitter (PBS), and a single photon detector (SPD). (b) Principle of the entanglement preparation stage. Photon-pairs in the diagonal state $\ket D \ket D$ enter the unbalanced MZI and exit in four possible states. If the photon-pair creation time remains unknown, the states $\ket H \ket H$ and $\ket V \ket V$ are indistinguishable in the time domain, and thus entanglement can be generated by post-selecting only when the two photons exit in the same time-bin~\cite{Martin_TB_2013}.}
\end{figure}

Based on the experimental setup depicted in \figurename~\ref{Fig_Exp}, we outline in the following the principles of the main building blocks.

\subsection{Photon-pair generation via type-0 spontaneous parametric down-conversion}

Light from a continuous-wave (CW) pump laser at 780.24\,nm is sent through a type-0 periodically poled lithium niobate waveguide (PPLN/W) in order to generate, via spontaneous parametric down-conversion (SPDC), paired photons around the degenerate wavelength of 1560.48\,nm. Using a vertically polarized (V) pump beam, the type-0 interaction permits to create vertically polarized twin photons, \textit{i.e.}, $\ket V_p \stackrel{type-0}{\longmapsto} \ket V_s \ket V_i$, where the indices $p$, $s$, and $i$ represent the pump, signal and idler modes, respectively. Utilizing the type-0 process presents two main advantages compared to the type-II interaction which produces cross-polarized photons, \textit{i.e.}, in the state $\ket H_p \stackrel{type-II}{\longmapsto} \ket H_s \ket V_i$, as detailed in~\cite{kaiserI_2012a} and references therein.
On one hand, the associated generation efficiency is at least 2 orders of magnitude higher. On the other hand, single photon bandwidth narrow filtering can readily be achieved using fiber filters, without problems with fiber birefringence.
However, in contrast to the type-II interaction, where polarization entanglement can be formed by simply splitting the paired photons, the price to pay for using the type-0 process is a more complex experimental arrangement~\cite{Kaiser_type0_2013}.

\subsection{Bandwidth filtering stage}

In principle, the bandwidth of the paired photons can adapted at will depending on the desired quantum application.
In our case, we work with fiber solutions only, compatible with all other elements of the source. By doing so, we avoid the losses due to in and out fiber coupling, yielding a higher long-term stability. After collection of the paired photons using a single mode fiber, we therefore take advantage of three filtering solutions based on telecom compatible components in order to demonstrate the versatility of our approach:
\begin{enumerate}[(a)]
\item a standard 100\,GHz-spacing dense wavelength division multiplexing (DWDM) filter (AC photonics) compatible with DWDM-QKD protocols~\cite{Yoshino_HSQKD_2012,kaiserI_2012a};
\item a 540\,MHz phase shifted fiber Bragg grating filter (PSFBG, from AOS GmbH) compatible with broad acceptance bandwidth quantum memories, based on, \textit{e.g.}, room temperature atomic vapors or ion doped crystals~\cite{reim_2010,clausen_2011,saglamyurek_2011};
\item a 25\,MHz PSFBG (Teraxion) compatible with the acceptance bandwidth of cold atom, and trapped ion based quantum memories~\cite{tanji_2009,piro_2011}.
\end{enumerate}

Moreover, to avoid any polarization dependent transmission, and therefore degradation of entanglement, with the narrowband PSFBG filters, those have to placed in front of the state preparation stage. The polarization state of the photons is adjusted beforehand using a fiber polarization controller (PC$_1$) (see more details below). 

\subsection{Polarization entanglement preparation stage}

To prepare the polarization entangled state, we employ an unbalanced MZI made of two fiber polarizing beam-splitters (f-PBS) connected by polarization maintaining fibers. This interferometer introduces a delay $\delta t$ between the two polarization modes H (short arm) and V (long arm). By sending the paired photons prepared in the diagonal (D) state $\ket D_s \ket D_i$ into this device, the exit states are of the form:
\begin{equation}
\label{Eq_outputMZI}
\ket D_s \ket D_i \stackrel{Prep.}{\mapsto} \frac{1}{2} \left[ \ket H_{s,e} \ket H_{i,e} + e^{i\phi/2} \ket H_{s,e} \ket V_{i,l}+ e^{i\phi/2}\ket V_{s,l} \ket H_{i,e}+ e^{i\phi}\ket V_{s,l} \ket V_{i,l}\right],
\end{equation}
where $e$ and $l$ refer to ``early'' and ``late'' time bins, respectively. Here, $\phi/2$ represents the phase difference between the short and long paths. Post-selecting only the cases where the two photons exit the interferometer simultaneously reduces the quantum state to $\ket{\psi}_{\rm post} = \frac{1}{\sqrt{2}} \left[ \ket H_{s,e} \ket H_{i,e}+  e^{ i\phi}\ket V_{s,l} \ket V_{i,l}\right]$.
This time-bin type post-selection constraints $\delta t$ to be greater than the coherence time of the photon ($\tau_p$) and the detector timing jitters ($\tau_d$). It is important to note that the labels $e$ and $l$ have physical meanings only if the photon-pair creation time is known. In the case of CW SPDC, the creation time uncertainty is directly given by the coherence time of the pump laser ($\tau_L$). In this way, using a laser with a coherence time much greater than $\delta t$ ensures a constant phase for the two contributions of interest, $\ket H_{s} \ket H_{i}$ and $\ket V_{s} \ket V_{i}$, and therefore allows obtaining polarization entangled states of the form:
\begin{equation}
\label{Eq_MaxES}
\ket{\psi}_{\rm post} = \frac{1}{\sqrt{2}} \left[ \ket H_{s} \ket H_{i}+  e^{ i\phi}\ket V_{s} \ket V_{i}\right],
\end{equation}
where $\phi$ represents the phase experienced by the two-photon contribution $\ket V_{s} \ket V_{i}$ in the long arm of the interferometer.
To summarize the operation principle described above, the source can produce, together with a proper post-selection, the maximally polarization entangled state of Eq.~\ref{Eq_MaxES} only if the conditions 
\begin{equation}
\label{Eq_cond}
\tau_L\gg \delta t \gg \tau_p +\tau_d
\end{equation}
are satisfied, as is the case in time-bin type experiments~\cite{Tittel_2001,Martin_TB_2013,Franson_Bell_1989}.

By controlling the phase difference between the arms of the MZI, and by adjusting the input polarization state of the two photons, it is possible to prepare any arbitrary superposition of the maximally entangled Bell states $\ket{\Phi^+}$ and $\ket{\Phi^-}$\footnote{There are four maximally entangled states referred to as Bell states~\cite{Tittel_2001}. They are defined as $\ket{\Phi^\pm} = \frac{1}{\sqrt{2}} \left[ \ket H_{s} \ket H_{i} \pm \ket V_{s} \ket V_{i} \right ]$ and $\ket{\Psi^\pm} = \frac{1}{\sqrt{2}} \left[ \ket H_{s} \ket V_{i} \pm \ket H_{s} \ket V_{i} \right ]$, and form a complete orthonormal basis.}. Suppose that the input of the preparation stage is the two-photon state $(\alpha \ket H + \beta \ket V) (\alpha \ket H + \beta \ket V)$, with the normalization $|\alpha|^2+|\beta|^2=1$. In this case we obtain, after appropriate post-selection of events in the time domain, non-maximally entangled states of the form $\ket{\psi}_{\rm post} = \frac{1}{N} \left[\alpha^2 \ket H_s \ket H_i + {e}^{ i\phi} \beta^2 \ket V_s \ket V_i\right]$, where $N = \sqrt{\alpha^4+\beta^4}$ is the normalization. This can be further generalized by simply adding a half-wave plate in the path of one of the two photons. This way, one can prepare non-maximally entangled states as a superposition of the four maximally entangled Bell states.

\section{Setup and characterization of the key elements}

\subsection{Frequency stabilization of the pump laser}

As laser source, we use a commercial tapered amplifier system (Toptica Photonics TApro) operating at a wavelength around 780.24\,nm in the CW regime. A small fraction of the laser light ($\rm \sim1\,mW$) is sent to a standard saturated absorption spectroscopy setup for frequency stabilization purpose. The stabilization is made on the $\ket{F=2} \rightarrow \ket{F'=2 \times 3}$ hyperfine crossover transition of the D$_2$ line of rubidium atoms ($^{87}$Rb). The Zeeman shifted resonance is modulated by $\sim$20\,kHz using a small electromagnet. The saturation absorption signal is then demodulated using a lock-in amplifier (EG\&G 5210), and fed forward to the laser diode current through a home-made PID controller. The speed of the stabilization system is sufficient in order to achieve a laser linewidth < 150\,kHz, over all the relevant time scales, which guarantees a coherence time $\tau_L$ > 3\,$\mu$s. We can therefore fulfill the remaining requirement of inequality~(\ref{Eq_cond}) even for long coherence time photons, \textit{i.e.}, up to $\tau_p =$20\,ns, as will be detailed below.

\subsection{Phase matching, efficiency, and optical characterization of the nonlinear integrated waveguide}

The photon-pair generator is a 4.5\,cm long, home-made, waveguide, integrated at the surface of a periodically poled lithium niobate substrate using the so-called soft proton exchange (SPE) technique~\cite{chanvillard_SPE_2000}. This permits obtaining low-loss ($\sim$0.2\,dB/cm) single-mode optical waveguide structures at telecom wavelengths, showing reasonably high mode confinement, via index variations on the order of 2$\times$10$^{-2}$. It also allows to maintain the integrity of the second-order non-linearity ($[\chi]^2$) of the lithium niobate substrate.
The desired quasi phase-matching condition (780.24\,nm $\mapsto$ 1560.48\,nm) for the type-0 SPDC process is obtained for a 7\,$\mu$m-width waveguide, a poling period of the substrate of 16.3\,$\mu$m, and for a temperature of 387\,K. Note that heating the sample also permits canceling the photorefractive effect induced by the coupled pump intensity in the waveguide.

The PPLN/W internal down-conversion probability was characterized in the single photon counting regime, as described in Ref.~\cite{Tanzilli_PPLNW_2002}, and found to be $\sim$4.8$\times$10$^{-6}$ generated pairs per injected pump photon. This outperforms photon-pair generators based on optical parametric oscillators~\cite{Wang2004,bao_2008} and bulk crystals~\cite{Fiorentino2008,piro_2009} by a few orders of magnitude, and is comparable to the best values reported to date for SPDC in type-0 PPLN/Ws~\cite{Tanzilli_PPLNW_2002,Halder_HighCoh_2008}. Such a high SPDC conversion probability can mainly be ascribed to the tight light confinement in the waveguide structure, since the efficiency grows quadratically with the optical power density~\cite{Tanzilli_Genesis_2012}.
Moreover, we also characterized the internal conversion efficiency in a more standard manner by using the opposite nonlinear process to SPDC, namely that of second harmonic generation. In this regime, when the quasi phase-matching condition is met, laser light at 1560\,nm is converted to 780\,nm with an internal efficiency of about 80\%/W/cm$^2$.

\begin{figure}[h!]
\center
\includegraphics[width=1\columnwidth]{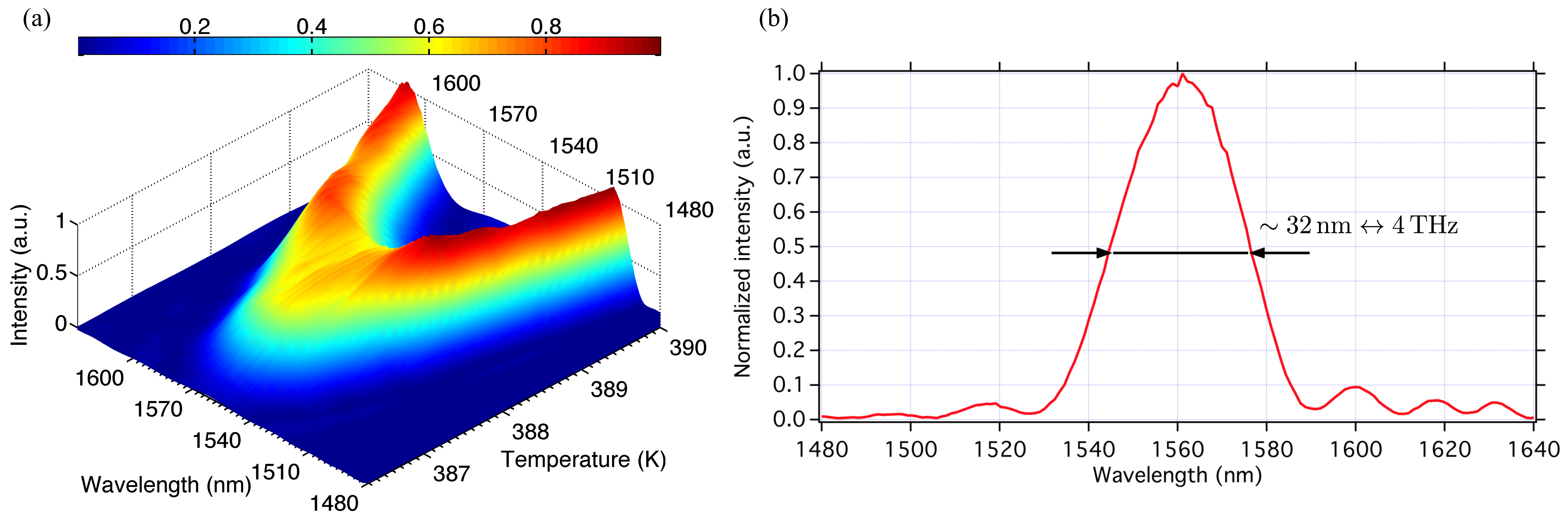}
\caption{\label{Fig_PDCemission}SPDC emission spectrum obtained from a PPLN/W having a poling period of 16.3\,$\mu$m, a width of 7\,$\mu$m, being pumped by a laser at 780.24\,nm. (a) Spectral density over the full phase matching curve as a function of temperature. The colorbar on top indicates the relative intensity. Changing the crystal temperature by 3\,K results in a wavelength tunability of more than 50\,nm for either the signal or the idler photon. (b) Degenerate spectrum around 1560\,nm, when the PPLN/W is heated to 387\,K. The shape of the curve is a sinc function, as expected from the SPDC process, and its FWHM spectral bandwidth is of 32\,nm, corresponding to 4\,THz.}
\end{figure}

The temperature dependent emission spectrum measured at the output of the PPLN/W is shown in \figurename~\ref{Fig_PDCemission}(a).
Signal and idler wavelengths can each be tuned over more than 50\,nm for a temperature change of 3\,K. In the following of this paper, the PPLN/W is operated at 387\,K, for which paired photons are emitted, as shown in \figurename~\ref{Fig_PDCemission}(b), at the degenerate wavelength of 1560.48\,nm, within a spectral bandwidth of 4\,THz ($\leftrightarrow32\rm \,nm$). Note that such a broad bandwidth is in general unsuitable for many quantum communication applications, since, \textit{e.g.}, QKD in a DWDM environment requires $50-200$\,GHz photonic bandwidths, while quantum memory protocols typically demand bandwidths reduced down to 10\,MHz$-$5\,GHz. However, 4\,THz is a broad enough bandwidth that can be further reduced so as to match a desired quantum application.

\subsection{Characterization of the transmission of the filters}

As mentioned previously, we employ three different fiber filters, \textit{i.e.}, a standard 100\,GHz DWDM, a 540\,MHz PSFBG, and a 25\,MHz PSFBG, to reduce the photonic spectral bandwidth from the 4\,THz initially generated by the PPLN/W. These filter bandwidths, together with the initial one, span more than 5 orders of magnitude.

\begin{figure}[h!]
\center
\includegraphics[width=1\columnwidth]{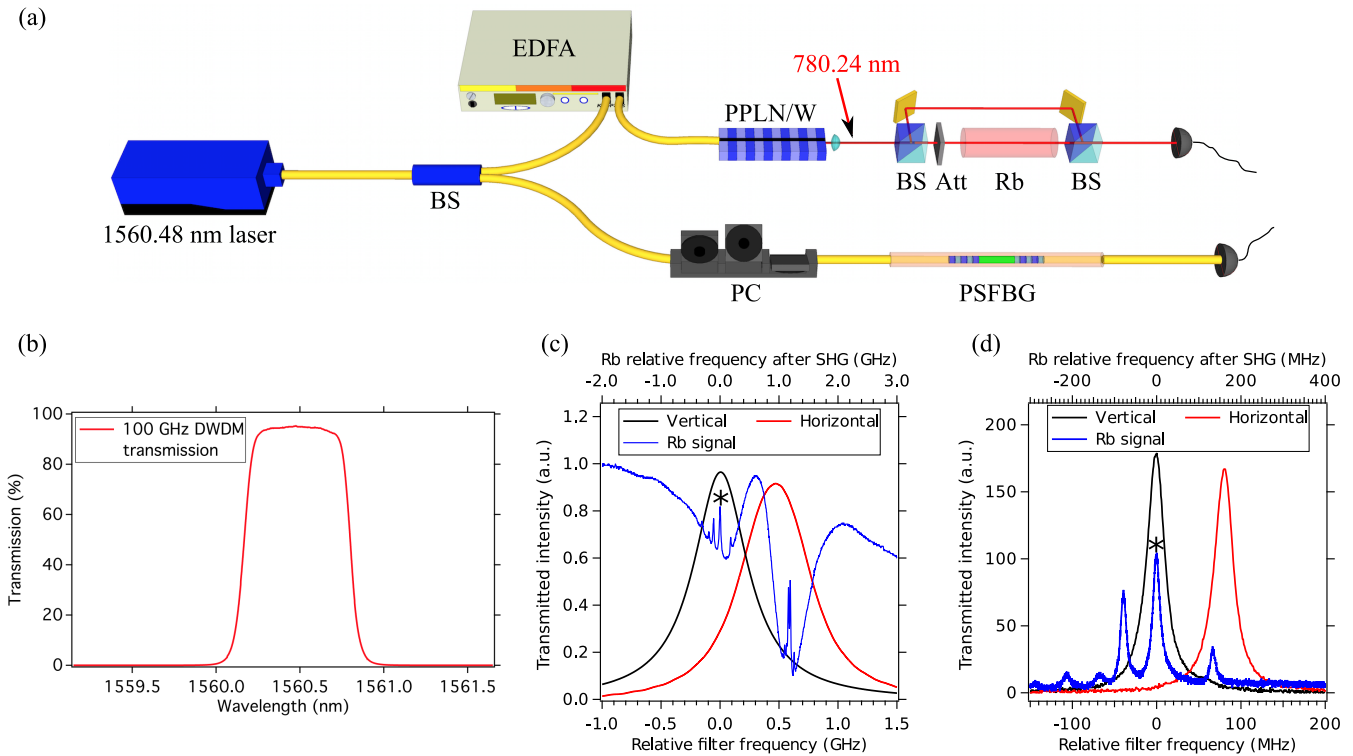}
\caption{\label{Fig_FBG}Characterization of the employed filters. (a) Experimental setup. A tunable laser, emitting around the wavelength of 1560.48\,nm, is sent through the filter under test and the polarization dependent transmission spectrum is recorded. In order to obtain a frequency reference in the case of the two PSFBG filters, the laser is amplified (EDFA) and frequency doubled in a PPLN/W, before being sent to a standard Rb saturated absorption spectroscopy setup. (b) Transmission spectrum of the 100\,GHz DWDM filter showing a flat-top shaped profile. (c) Transmission spectrum of the 540\,MHz filter. (d) Transmission spectrum of the 25\,MHz filter. The two latter filters show lorentzian transmission profiles, which have a polarization dependence caused by fiber birefringence.}
\end{figure}

As depicted in \figurename~\ref{Fig_FBG}(a), we properly characterize and set the three filters at the desired wavelength (1560.48\,nm). Notably, the two PSFBG filters are configured using laser spectroscopy. Fort this, we inject light from a tunable diode laser (Toptica Photonics DL100 pro design), followed by a polarization controller (PC), and we monitor the transmitted power at the output.
To obtain an accurate frequency reference, a fraction of the laser power is amplified using an erbium doped fiber amplifier (EDFA) and then frequency doubled in another home-made PPLN/W. For $\sim100\,\rm mW$ of power at 1560.48\,nm sent in the dedicated PPLN/W, we obtain $\sim 3\,\rm mW$ at 780.24\,nm at its output. This is enough to perform saturated absorption spectroscopy in a Rb cell. The obtained hyperfine structure absorption lines here serve as accurate absolute frequency references when scanning the laser.

As depicted in \figurename~\ref{Fig_FBG}(b), the 100\,GHz DWDM filter shows a flat-top transmission bandwidth of 80\,GHz. In \figurename~\ref{Fig_FBG}(c), we show the transmission of the 540\,MHz PSFBG, which has lorentzian transmission bandwidths of 540\,MHz and 580\,MHz for vertically and horizontally polarized light, respectively. Due to fiber birefringence, the modes are separated by 480\,MHz. The measurements for the 25\,MHz PSFBG are given in \figurename{~\ref{Fig_FBG}}(d), with lorentzian transmission bandwidths of 25\,MHz and 28\,MHz for vertically and horizontally polarized light, respectively, and a mode separation of 80\,MHz. Note that a consequence of the polarization dependent transmission behavior is that these PSFBGs cannot be directly applied to polarization entangled paired photons, as the frequency separation between horizontally and vertically polarized photons would reduce the entangled state to a product state.

The central frequency (wavelength) of both PSFBG filters can be finely tuned by adjusting the temperature. Experimentally, the 540\,MHz and the 25\,MHz filters present respectively a tunability of $\sim$1\,GHz/K and $\sim$200\,MHz/K, such that we can achieve about 50\,GHz ($\leftrightarrow$ 400\,pm) and 10\,GHz ($\leftrightarrow$ 80\,pm) over a temperature range between 30$^\circ$C and 80$^\circ$C. As a side effect, the temperature of the filters needs to be stabilized in order to avoid any central wavelength drift.
In our case, this is done by isolating the filters hermetically from ambient air and by employing standard PID-type temperature controllers. Using such systems, we can set the transmission peak of both filters for vertically polarized light at exactly twice the wavelength of the Rb $\ket{F=2} \rightarrow \ket{F'=2 \times 3}$ hyperfine transition, against which the pump laser is stabilized (indicated by a star in the graphs of \figurename{~\ref{Fig_FBG}}(b) and (c)). The achieved temperature stability is 10\,mK, corresponding to frequency stabilities of $\sim 10\,\rm MHz$ and $\sim 2\,\rm MHz$ for the 540\,MHz and the 25\,MHz filters, respectively.

Note that these PSFBG filters represent simple solutions compared to standard optical cavities, concerning both practicality and reliability. For instance, stabilizing a cavity to better than 2\,MHz requires a $\sim$1\,mK temperature stability, as well as a stringent isolation to ensure a constant pressure and a limited sensibility to acoustic vibrations. This is actually unnecessary with our fiber filters. Moreover, tuning the central frequency of a cavity necessitates the use of a piezo-electric transducer place on one of the cavity mirrors. This, in our case, can simply be achieved using basic temperature control. A good example of a narrow filtering made with an optical cavity can be found in reference~\cite{piro_2009}.

\subsection{Polarization entanglement preparation stage and coincidence histogram}

As described above, after paired photons have been generated in the PPLN/W and subsequently filtered in bandwidth (see \figurename~\ref{Fig_Exp}), we take advantage of an unbalanced MZI to engineer the polarization bi-photon state. When two diagonally polarized photons are injected in the device, they emerge in the state of Eq.~\ref{Eq_outputMZI}, which can be further reduced to the maximally polarization entangled state of Eq.~\ref{Eq_MaxES} by post-selection in the time domain.
A proper post-selection can only be achieved provided the different contributions to the state of Eq.~\ref{Eq_outputMZI} are made temporally distinguishable.
In other words, the delay $\delta t$ between the two arms of the interferometer has to be much greater than the coherence time $\tau_p$ of the photons, as stated by inequality~\ref{Eq_cond}.
For this reason, we have chosen a fiber path length difference of 18\,m, corresponding to a propagation time difference of 76\,ns, enough to obtain a fidelity greater than 99\% with spectral bandwidths as narrow as 19\,MHz (coherence time of 20.5\,ns).

\begin{figure}[h!]
\center
\includegraphics[width=0.75\columnwidth]{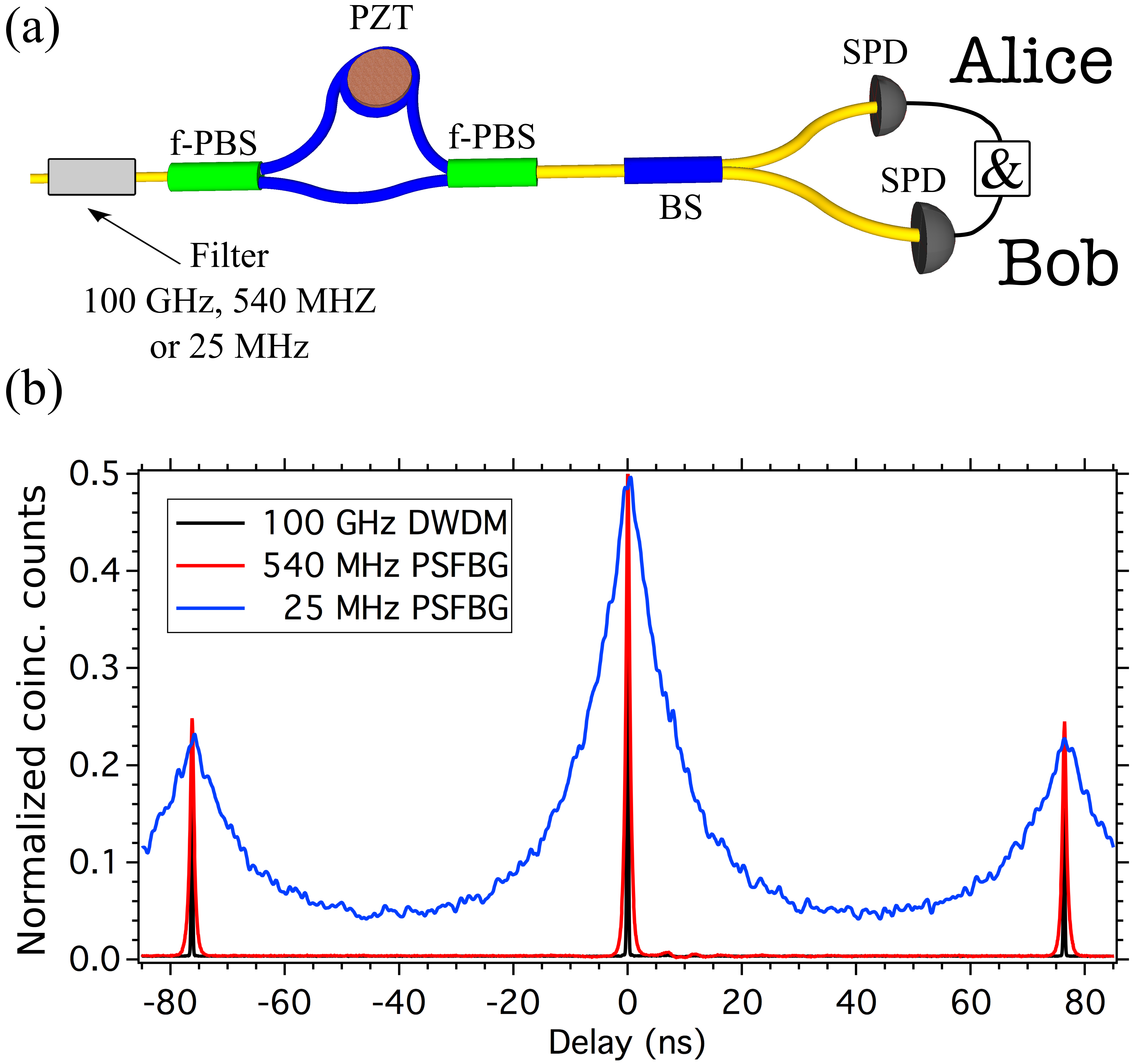}
\caption{\label{Fig_TAC}Characterization of the polarization entanglement preparation stage, in combination with the three filters, for a diagonally polarized two-photon input state. (a) Experimental setup. At the output of the preparation stage, the photons are separated by a BS and subsequently detected. The arrival time correlations are recorded using a coincidence counter (\&). (b) Coincidence histograms. Three well separated peaks are obtained for all the three filters. The two outer ones, located at $\pm$76\,ns, correspond to cross polarized contributions. The central peak at zero relative time delay contains the desired contributions $\ket H \ket H$ and $\ket V \ket V$.}
\end{figure}

We show in \figurename~\ref{Fig_TAC}(a) the experimental configuration employed for obtaining arrival time coincidence histograms at the output of the interferometer.
A beam-splitter (BS) is used to separate and distribute the photons to Alice and Bob. Each user employs a free running indium-gallium-arsenide (InGaAs) avalanche photodiode (IDQ-220) as single photon detector (SPD). Both devices feature 20\% detection efficiencies and dark count probabilities on the order of 10$^{-6}$/ns. The SPD outputs are connected to a coincidence measurement apparatus (\&).

As shown in \figurename~\ref{Fig_TAC}(b), we observe three distinct coincidence peaks for all three filters. The side peaks contain the cross-polarized photon-pair contributions, while the central peak is twice as high and contains the desired contributions $\ket H \ket H$ and $ \ket V \ket V$. Selecting only the events at zero relative delay generates the desired entangled state $\ket{\psi}_{\rm post}$ of Eq.~\ref{Eq_MaxES}.
For the 100\,GHz DWDM filter, we expect a coherence time $\sim 5$\,ps smaller than the timing jitters of the detectors. Consequently the width of the corresponding coincidence peaks (dark curve in \figurename~\ref{Fig_TAC}) is mainly given by the convolution of the detectors' timing jitters (230\,ps for each detector). Having the 540\,MHz or the 25\,MHz filter in place leads to a broadening of the coincidence peaks of 800$\pm$20\,ps (red curve in \figurename~\ref{Fig_TAC}) and 15.6$\pm$0.7\,ns (blue curve in \figurename~\ref{Fig_TAC}), respectively, which reflects the increased coherence time of the photons. These values are in good agreement with the corresponding filter bandwidths shown in \figurename{~\ref{Fig_FBG}}(b) and (c).

\subsection{Phase stabilization of the polarization entanglement preparation stage}

In order to demonstrate high quality polarization entanglement via the violation of the Bell inequalities (see below), the phase relation between the contributions $\ket H \ket H$ and $\ket V \ket V$ to the desired entangled state needs to be stable throughout the measurement time~\cite{martin_2012}. Our calculations showed that phase fluctuations on the order of $\Delta \phi \approx \frac{2\,\pi}{50}$ cause a degradation of the entanglement state fidelity of about 0.5\%.
In our MZI fiber configuration, phase variations are mainly due to refractive index fluctuations induced by temperature instabilities~\cite{leviton_2008}. The temperature phase dependency is approximately given by $\frac{\Delta \phi}{\Delta T} \approx 10^3\rm \,rad \,K^{-1}$. This means that achieving a phase stability better than $\frac{2\,\pi}{50}$ would require a temperature stability of about 30\,$\mu$K, which is technically demanding as the MZI would have to be placed in a vacuum chamber.

\begin{figure}[h!]
\center
\includegraphics[width=0.75\columnwidth]{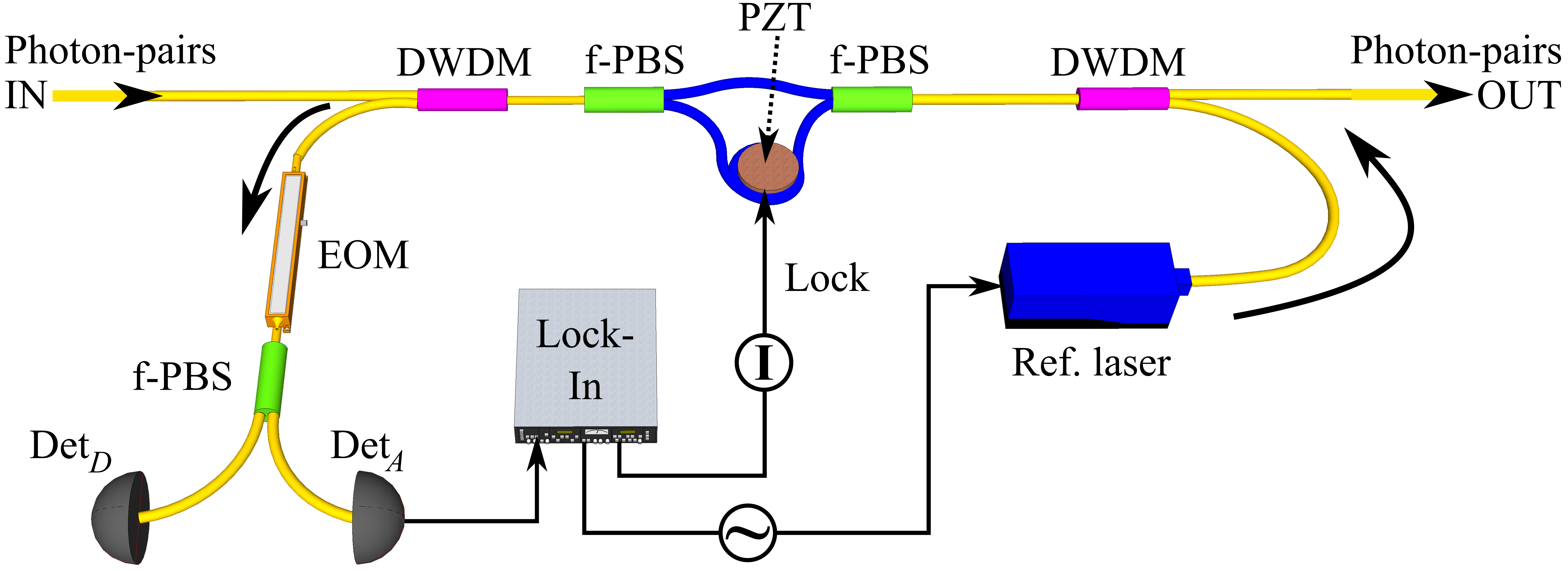}
\caption{\label{Fig_PhaseStabSetup}Phase stabilization system of the interferometric entanglement preparation stage. The phase accumulated by a reference laser is measured and a correction signal is forwarded to a PZT. The phase can be accurately tuned to any desired value by adding an offset phase with a phase modulator.}
\end{figure}

To avoid the latter constraint, we opted for an active stabilization scheme. As depicted in \figurename~\ref{Fig_PhaseStabSetup}, light from a reference diode laser (Toptica Photonics DL100 pro design), diagonally polarized and emitted at 1558.6\,nm, is injected in the interferometer via a telecom DWDM in the backward direction compared to that of the paired photons. Note that this component is a 200\,GHz-spacing DWDM allowing multiplexing and demultiplexing of the standard ITU\footnote{International Telecommunication Union.} telecommunication channels N$^\circ$21 and 23.
At the output of the MZI device, the laser light is separated from the photons of interest via another DWDM and sent to an electro-optical phase modulator.
The corresponding polarization state can therefore be written as $\ket{\psi}_{\rm r} = \frac{1}{\sqrt{2}} \left( \ket{H}+ e^{i(\phi_{\rm r} + \phi_{\rm e})}\,\ket{V} \right)$, where $\phi_{\rm r}$ is the relative phase difference between the short (horizontal) and long (vertical) arms of the MZI, and $\phi_{\rm e}$ a controllable phase added by the modulator to the vertical contribution. 
Finally, a fiber PBS is used to project the light into the $\{\ket D, \ket A \}$ basis, in which one obtains $\ket{\psi}_{\rm r} =  \sin \left( \frac{\phi_{\rm r} + \phi_{\rm e}}{2} \right) \ket{D}+ \cos \left( \frac{\phi_{\rm r} + \phi_{\rm e}}{2} \right) \ket{A}$. Consequently, measuring the light intensity at detector Det$_A$ allows us to infer the phase term $\left( \phi_{\rm r} + \phi_{\rm e} \right) {\rm mod} \, 2\pi$.

Our phase stabilization scheme is based on a standard feedback loop, in which the reference wavelength is modulated via the current controller of the laser diode. To obtain the error signal, a lock-in amplifier demodulates the detector signal. The phase correction is applied to a home-made piezoelectric (PZT) fiber stretcher acting on the long arm of the MZI via an integrator circuit. Our system holds the phase term $\phi_{\rm r} + \phi_{\rm e}$ constant at zero or $\pi$. The phase inside the interferometer, $\phi_{\rm r}$, is then tuned by adjusting $\phi_{\rm e}$. Our system is therefore capable of accurately setting the phase to any desired value over short time scales, \textit{i.e.} on the order of 1\,ms (mainly limited by the response time of the PZT).

Moreover, to ensure a long term stability, the 1558.6\,nm reference laser is stabilized in a cavity, after frequency doubling to 779.3\,nm. The cavity is itself stabilized upon that of the 780.24\,nm pump laser (see above), which means that the reference laser is stabilized thanks to  a transfer-cavity lock configuration. The overall system achieves an overall phase stability better than $\frac{\pi}{100}$ and works reliably over time scales of several days.

\section{Entanglement analysis}

In this section, we first demonstrate our capability to coherently tune the phase of the preparation interferometer. Thereafter, we set the phase to the desired value of $\pi$ in order to create the entangled state $\ket{\Phi^-}$, and we perform entanglement analysis via the violation of the Bell inequalities.

\subsection{Phase tuning demonstration and long-distance distribution of the paired photons}

\begin{figure}[h!]
\center
\includegraphics[width=0.75\columnwidth]{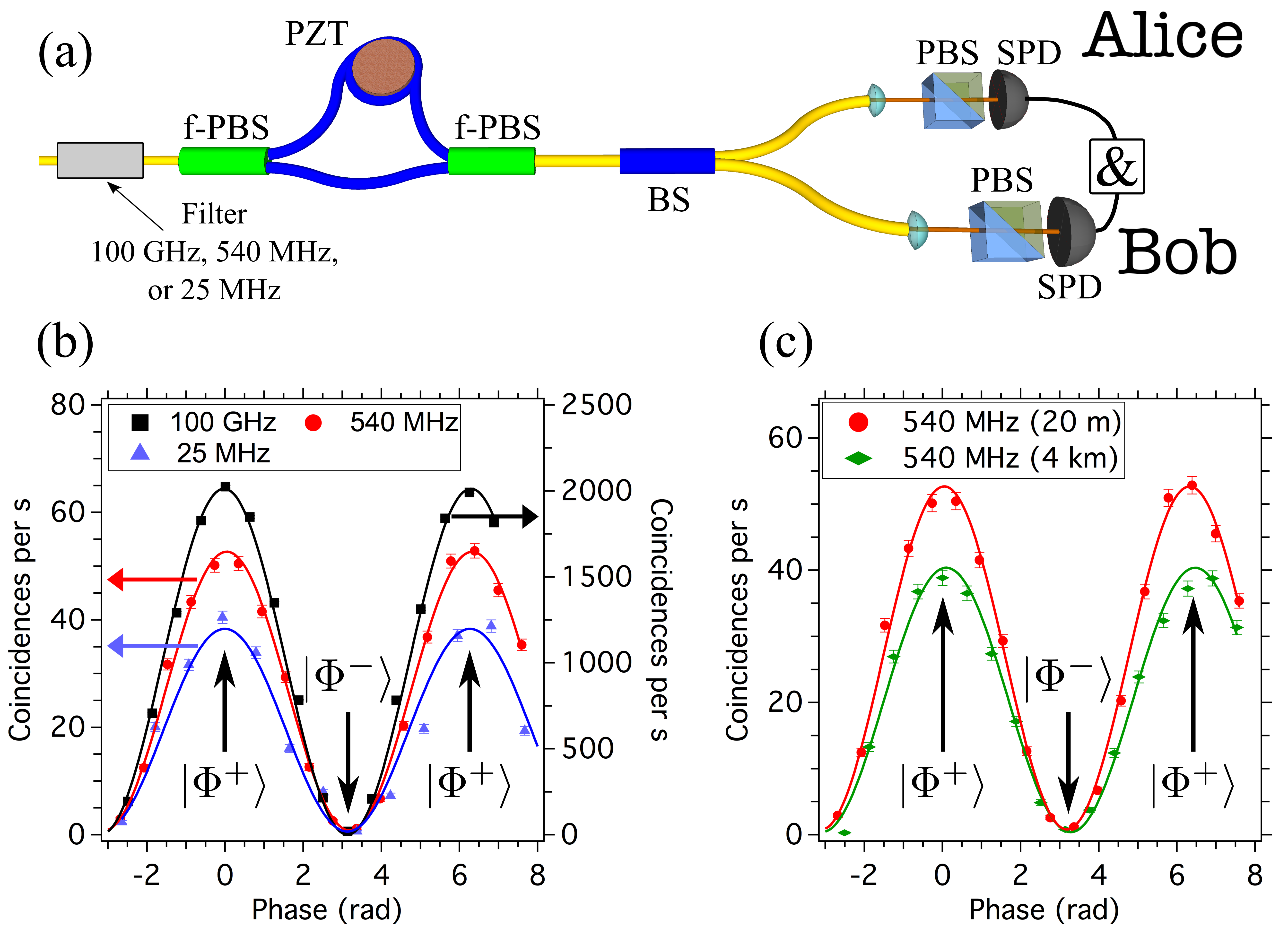}
\caption{\label{Phase}Demonstration of phase control capabilities. (a) Experimental setup. Alice and Bob both set their polarization state analyzers at 45$^{\circ}$, in order to project the entangled state on the phase sensitive $\{ \ket D, \ket A \}$ basis. The coincidence rate between the two detectors is recorded as a function of the phase set in the fiber preparation stage. (b) Results obtained for photon-pair distribution over 20\,m of standard optical fiber: for all three filters, near perfect sinusoidal coincidence rate oscillations are observed, underlining the high phase stability achieved with our approach. (c) Results obtained for the 540\,MHz filter when distributing the entangled photons over 20\,m and 4\,km of standard optical fiber, respectively. Again, high-visibility interference patterns are observed. Note that for all results presented here, noise contributions induced by accidental coincidences have been subtracted.}
\end{figure}

The experimental setup used to demonstrate the phase tunability is shown in \figurename~\ref{Phase}(a). The photon-pairs, at the output of the interferometric preparation stage, are expected to be prepared, together with a proper time-bin post-selection, in the state $\ket{\psi}_{\rm post}$ (see Eq.~\ref{Eq_MaxES} and related discussion).
The paired photons are then separated at a beam-splitter and sent to Alice and Bob. They both project their respective photons into the the phase sensitive diagonal/antidiagonal basis $\{ \ket D, \ket A \}$ using a polarization state analyzer made of a bulk polarizing beam-splitter (PBS) oriented at $45^{\circ}$ and an InGaAs avalanche SPD. In this situation, the coincidence rate, $R_{\rm c}$, recorded between the two users is given by
\begin{equation}\label{Eq_PhaseFormula}
R_{\rm c} = \frac{1}{2}\left( 1 - V \cos {\phi} \right),
\end{equation}
where $\phi$ is the phase imprinted on the entangled state by the interferometric preparation stage, and $V$ represents the fringe visibility of the interference pattern.

In \figurename~\ref{Phase}(b), we present the coincidence rate recorded as a function of the phase set in the preparation stage for all three filters. For these measurements, the phase is tuned between $-\pi$ and $5\pi/2$, using our stabilization system.
The three curves follow the relation described by equation~\ref{Eq_PhaseFormula}, and sinusoidal fits permit to infer the associated interference pattern visibilities. The net visibilities are $99.9\pm 1.2\%$, $99.4\pm 1.5\%$, and $97\pm 2\%$, for the 100\,GHz (squares and dark line), 540\,MHz (circles and red line), and 25\,MHz (triangles and blue line) filters, respectively. Net visibilities mean that accidental coincidences induced by the detector dark counts have been subtracted. Without noise subtraction, the raw visibilities are $99.9\pm 1.2\%$, $97.1\pm 1.2\%$, and $88\pm 2\%$, respectively. The last value is smaller than the others, owing to a degraded signal-to-noise ratio associated with the 25\,MHz.

Moreover, \figurename~\ref{Phase}(c) shows the results obtained with the 540\,MHz filter when the paired photons are distributed over 4\,km of standard telecom optical fiber, and its direct comparison to a 20\,m fiber distribution scenario. The obtained net and raw visibilities are $99\pm2\%$ and $96\pm2\%$, respectively. The reductions in coincidence rate and raw visibility are ascribed to additional propagation losses of about 1.2\,dB, which again causes a degradation of the signal-to-noise ratio.

To conclude this phase control and tunability study, it is worth mentioning that our preparation system works perfectly as it permits setting the phase of the entangled state to any desired value on short time scales (< 1\,ms). The curves presented in \figurename~\ref{Phase} therefore stand as a proof that any maximally entangled state of the form $\ket{\psi} = \frac{1}{\sqrt{2}} \left( \ket{H} \ket{H}+ {\rm e}^{{\rm i}\phi}\,\ket{V} \ket{V}\right)$ can be generated at will. For the following study, we now set $\phi = \pi$ in order to generate the state $\ket{\Phi^-}$, on which entanglement analysis is performed.

\subsection{Entanglement analysis for the $\ket{\Phi^-}$ state}

\begin{figure}[h!]
\center
\includegraphics[width=1\columnwidth]{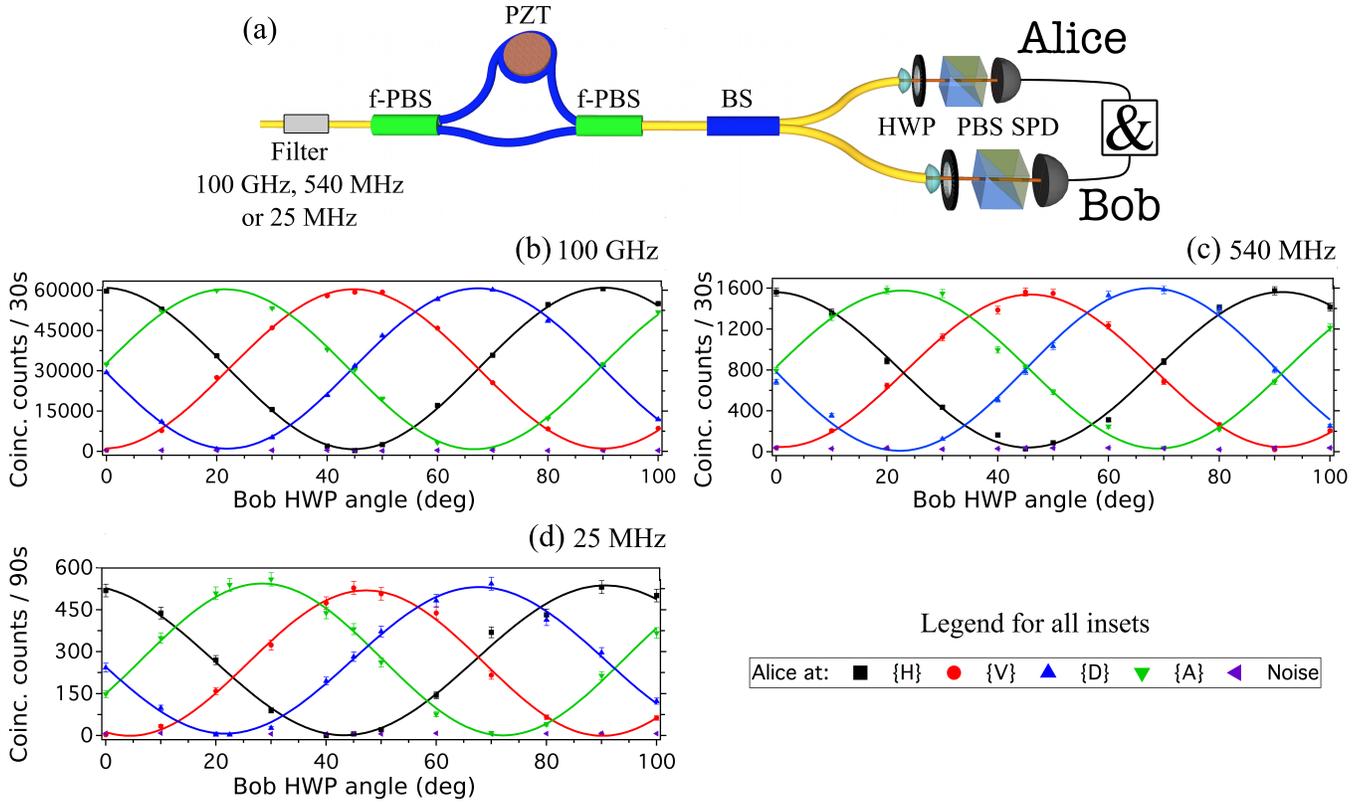}
\caption{\label{Fig_Bell}Demonstration of entanglement quality via the violation of the Bell inequalities. (a) Experimental setup. Both users utilize a polarization state analyzer consisting of a half-wave plate (HWP), a PBS, and a SPD. The coincidence rate correlations for the standard polarization settings are recorded. For all three filters, DWDM (b), 540\,MHz PSFBG (c) and 25\,MHz PSFBG (d), near perfect visibility two-photon interference patterns are obtained for all polarization settings. This underlines the high quality entanglement generated with out approach.}
\end{figure}

In the following, we employ, among other possibilities~\cite{Altepeter_experimental_2005}, the violation of the Bell inequalities as an entanglement witness~\cite{bell_1964,clauser_1969}. To this end, we consider the standard experimental configuration shown in \figurename~\ref{Fig_Bell}(a), in which the users each employ a polarization state analyzer, comprising a half-wave plate (HWP), a PBS, and an SPD. As was the case for the phase control study, the two detectors are connected to a coincidence measurement apparatus (\&).

The measurements are made in the following manner. Alice projects her photons consecutively along the horizontal ($\ket H$), vertical ($\ket V$), diagonal ($\ket D$) and anti-diagonal ($\ket A$) directions. For each of the four settings, we record the coincidence rates as a function of Bob's HWP angle. The experimental results corresponding to the three filters are shown in \figurename~\ref{Fig_Bell}(b), (c) and (d). For each filter and for all four settings chosen at Alice's side, sinusoidal coincidence rate oscillations are observed, underlining the rotational invariance of the generated entangled state. Sinusoidal fits permit to infer the associated interference pattern visibilities. We obtain average raw visibilities (including noise contributions) of $99.6\pm1.3\%$, $97.1\pm0.9\%$ and $99\pm3\%$ for the 100\,GHz, 540\,MHz, and 25\,MHz filters, respectively. These values are clearly above the non-local threshold of 71\% for two-photon experiments~\cite{Tittel_BellGeneral_1999}. Note that in order to increase the visibility for the 25\,MHz filter, two superconducting detectors (Scontel TCOPRS-001) have been used instead of the two InGaAs avalanche photodiodes that were in place for all the previous measurements. These superconducting devices both feature 7\% detection efficiencies and less than 10 dark counts per second. This permits us to increase the signal-to-noise ratio at the price of lower coincidence counts.

Besides the obtained visibility, the Bell parameter is another quantitative measure for evaluating the quality of entanglement. We obtain raw Bell parameters $S_{\rm raw}$ of $2.82\pm0.01$, $2.80\pm0.02$ and $2.82\pm0.02$ for the three employed filters, meaning that the Bell inequalities are violated by at least 40 standard deviations~\cite{bell_1964,clauser_1969}. This underlines the high quality of the entangled states that can be generated with our scheme~\cite{Kaiser_type0_2013}.

\section{Brightness and internal losses of the source}

In addition to high entanglement qualities, the brightness of the photon-pair generator is one of its key features, since it provides the rate at which a quantum communication link can be operated. Typically, the brightness is defined as the number of paired photons available in one spatial mode, per of integration time, pump power, and spectral bandwidth. Achieving a high brightness requires the use of a high-efficiency generator associated with low propagation losses in the overall setup. In the following, we first detail the two latter figures, before giving the rate of entangled photon-pairs available at the output of the source.

On one hand, the internal down-conversion probability of the type-0 PPLN waveguide obtained over the full emission spectral bandwidth of 4\,THz, shown in \figurename~\ref{Fig_PDCemission}(b), corresponds to a brightness $B_{\rm full} \approx 2400\,\rm \,pairs \cdot (s \cdot mW \cdot MHz)^{-1}$. However, all the three employed filters provide much narrower bandwidths than that of the full spectrum, such that the peak intensity within these bandwidths can be considered constant with an associated brightness $B_{\rm top} \approx 3600\,\rm \,pairs \cdot (s \cdot mW \cdot MHz)^{-1}$. As for all generators based on spontaneous processes, the main limitation comes from multiple pair emission issues, generally dominated, at weak pump powers, by double pair contributions. Those are due to the emission statistics that can be either poissonian or thermal, depending on both the coherence time of the photons and the time acquisition window~\cite{Tapster1998,Monteiro_Pure_2014}. In other words, operating the source at high pump powers would lead to non negligible double pair contributions and cause a severe degradation of the entangled state fidelity~\cite{Scarani_2pairs_04}. For instance, working with a probability for single photon-pair emission equal to 0.01 per detection time window, generally defined as the maximum between the timing jitter of the detectors and the coherence time of the photons, leads to a decrease of the relative entangled state fidelity of 1\%. Note that with our non-linear waveguide, a laser pump power of 7\,mW is enough to achieve such a probability value per coherence time.

On the other hand, to compute the overall source brightness, we need to consider all losses from the output of the photon-pair generator to the output of the source itself, \textit{i.e.}, the two outputs of the fiber beam-splitter shown in \figurename~\ref{Fig_Exp}. We measure the single photon propagation losses between the output of the crystal and the input of the beam-splitter by replacing the pump laser by one emitting at 1560.48\,nm. We obtain 4.5\,dB, 5.2\,dB, and 5.7\,dB for the 100\,GHz, 540\,MHz and 25\,MHz filters, respectively. However, when considering the propagation of the paired photons to the output of the source, additional losses have to be taken into account. We find the following values:
\begin{itemize}
\item 3\,dB due to the lorentzian transmission profiles of the PSFBG filters, which are applied to energy correlated paired photons produced in the CW pump regime. Note that this is not the case for the DWDM filter, since it shows a flat-top transmission profile;
\item 3\,dB due to the time-bin structure associated with the preparation stage, since only half of the generated photon-pairs are actually prepared in a maximally polarization entangled state;
\item 3\,dB introduced by a non-deterministic pair separation at the output beam-splitter.
\end{itemize}
The overall losses experienced by the paired photons are therefore 15\,dB, 19.4\,dB and 20.4\,dB, respectively for the three filter configurations.

\section{Summary of the source performances and comparison to other realizations}

Combining the generator brightness and the internal losses of the full setup leads to available photon-pair rates of about 9$\times$10$^6$, 22000, and $\sim$800 per second and mW of pump power, respectively for the three filter configurations.
A summary of the relevant figures of merit for the source, in combination with the three different filters, is presented in table~\ref{Table_Source_type0_sum}.

\begin{table*}[!h]
\begin{center}
\begin{tabular}{| l || c | c | c |} \hline
Bandwidth [MHz] & 80$\times 10^{3\dagger}$ & 540$^\dagger$ & 25$^\ddagger$ \\ \hline
Photon coherence time [ns] & 5.5$\times 10^{-3}$ & 0.8 & 15.6 \\ \hline
Injected pump power [mW] & 0.02 & 0.4 & 7 \\ \hline
Photon-pair losses [dB] & 15 & 19.4 & 20.4 \\ \hline
Available output rate [(s$\times$mW)$^{-1}$] & 9$\times$10$^6$ & 22$\times$10$^3$ & 800 \\ \hline
Detected pair rate [s$^{-1}$] & 2000 & 50 & 6 \\ \hline
Interference raw visibility (\%) & 99.6\,$\pm1.3$ &  97.1\,$\pm0.9$  & 99\,$\pm$3 \\ \hline
\end{tabular}
\caption{\label{Table_Source_type0_sum}Summary of the experimental results carried out with the experimental setup shown in \figurename~\ref{Fig_Exp} for the three different filters.
Note that the imperfections of the detectors provide limitations on data acquisition, since low efficiencies are associated with lower pair detection rates, and dark counts are one of the main limiting factor for achieving high raw visibility values~\cite{Sekatski2012}. Moreover, for both the 100\,GHz DWDM and the 540\,MHz PSFBG filters, the dead-times of the detectors induce an inherent reduction of the detected pair rate.
$^\dagger$\,For these measurements, standard InGaAs avalanche photodiode SPDs have been used~\cite{hadfield_2009}.
$^\ddagger$\,For this particular measurement, superconducting SPDs have been employed in order to reduce accidental coincidence events~\cite{hadfield_2009}.}
\end{center}
\end{table*}

Finally, table~\ref{Table_Source_type0_comp} presents a detailed comparison of our source performances for the three utilized filters to other pertinent and recent polarization entangled photon-pair sources reported in the literature. The purpose of this table is definitely not to be exhaustive, but rather to give an overview of different types of realizations, sorted by group, generator and arrangement types, spectral bandwidth of interest ($\Delta \nu$), emitted wavelength ($\lambda$), normalized brightness (B), and quality of entanglement (visibility $V$).

\begin{table*}[!h]
\begin{center}
\begin{tabular}{|p{3.4cm}|p{2.8cm}|p{2cm}|p{1.7cm}|p{1.5cm}|c|} \hline \hline
Group & Generator &	\centering $\Delta \nu$	(MHz) & \centering $\lambda$	(nm) &	\centering B$^{(\star)}$ &  $V$\\
 \hline
\cite{Kuklewicz_2006}~Cambridge (2006) & KTP OPO & \centering 22$^{(a)}$ & \centering 795 & \centering 0.7 & 77\% \\
\cite{bao_2008}~Hefei (2008) & PPKTP OPO & \centering 9.6 &	\centering 780	 &	\centering 6 & 97\% \\
\cite{piro_2009}~Barcelona (2009) & PPKTP & \centering 22 & \centering 854 & \centering 3 & 98\% \\
\cite{dousse_2010}~Paris (2010) & Quantum dot & \centering NA$^{(b)}$ & \centering $\sim$900 & \centering NA$^{(c)}$ & 36\%$^{(d)}$ \\
\cite{yan_2011}~Hong Kong (2011) & Rb atoms & \centering 6 & \centering 780 & \centering 0.5 & 89\% \\
\cite{kaiserI_2012a}~Nice (2012) & type-II PPLN/W & \centering 80$\times$10$^3$ & \centering 1540 & \centering 20 & 97\% \\
\cite{Kaiser_type0_2013}~Nice (this work) & type-0 PPLN/W & \centering 80$\times$10$^3$ & \centering 1560 & \centering 114$^{(e)}$ & 99\% \\
\cite{Kaiser_type0_2013}~Nice (this work) & type-0 PPLN/W & \centering 540 & \centering 1560 & \centering 41$^{(e)}$ & 97\% \\
\cite{Kaiser_type0_2013}~Nice (this work) & type-0 PPLN/W & \centering 25 & \centering 1560 & \centering 33$^{(e)}$ & 99\% \\ \hline
\hline
\end{tabular}
\caption{\label{Table_Source_type0_comp}Comparison of recent narrowband entangled photon-pair sources sorted by group, generator and arrangement types, spectral bandwidth ($\Delta \nu$), emitted wavelength ($\lambda$), normalized brightness (B), and entanglement visibility ($V$). $^{(\star)}$B is calculated as mentioned in the above Section, \textit{i.e.} the number of entangled photons available after both the filtering stage and per single spatial mode (Pairs/($\rm s \cdot mW \cdot MHz$)). $^{(a)}$In Ref.~\cite{Kuklewicz_2006}, 22\,MHz corresponds to the spectral linewidth of one longitudinal mode of the OPO cavity. $^{(b)}$The actual photonic bandwitdth is not provided in Ref.~\cite{dousse_2010} as the produced photons are not Fourier transform limited. $^{(c)}$In Ref.~\cite{dousse_2010}, the authors report an emission probability of 0.12 per excitation laser pulse. $^{(d)}$In Ref.~\cite{dousse_2010}, the authors report a maximum fidelity to the closest entangled state limited to 0.68. $^{(e)}$The reported B values for the three filter configurations do not correspond to that reported in Ref.~\cite{Kaiser_type0_2013}, in which the propagation losses from the output facet of the generator to the distribution fibers have not been taken into account.}
\end{center}
\end{table*}

\newpage

\section{Potential improvements of the source}

The pair rate could be improved considerably by reducing the internal propagation losses of the source, notably in the case of the narrowest employed filters.
First, the lorentzian PSFBG filters (540\,MHz and 25\,MHz) could be replaced by flat-top ones, as for the 100\,GHz DWDM filter. This would, in this case, lead to a pair loss reduction of 3\,dB. 
Secondly, the propagation losses could also be reduced by splicing all the fiber patch cords instead of using fiber connectors. This would lead to approximately a 2\,dB pair loss reduction. Thirdly, using a tapered waveguide structure at the output of the non-linear waveguide generator would yield another 1\,dB of loss reduction, due to an adiabatic spatial mode adaptation between the waveguide and the collection fiber~\cite{castaldini_2009}. 
Finally, one could address the non-deterministic separation of the paired photons by utilizing a fiber coupled cavity able to transmit two peaks on both sides of the wavelength degeneracy (1560.48\,nm). For instance, a free spectral range of 100\,GHz combined with a transmission bandwidth of 25\,MHz can be achieved by using a 1.2\,mm-long cavity, finesse 4000. Despite the difficulty of building such a high finesse cavity, this would permit obtaining pairs of entangled photons easily separable using a simple DWDM placed after the interferometric preparation stage, providing a 3\,dB pair loss reduction.
Combining some of the above mentioned strategies could lead to an average pair loss reduction on the order of 6\,dB, at the price of an increased setup complexity.

\section{Conclusion}

We have demonstrated a versatile and efficient scheme for generating polarization entangled photon-pairs at telecom wavelengths. The versatility lies in both the spectral properties of the photons and in the generated quantum state. For instance, the wavelength of the emitted single photons can easily be adjusted over more than 50\,nm, using basic temperature control of the photon-pair generator. We have also shown near perfect entanglement qualities over photonic spectral bandwidths ranging from 25\,MHZ to 100\,GHz, thus spanning more than four orders of magnitude. Our scheme is readily extendable to the 4\,THz spectral bandwidth initially available at the output of the photon-pair generator, and therefore to five orders of magnitude. Moreover, the active stabilization system of the preparation stage allows us to set and to further tune the phase relation between the two contributions, $\ket{HH}$ and $\ket{VV}$, to the entangled state. The weight of the contributions, and the contributions themselves, can be easily chosen by means of basic polarization controls in front of and after the preparation stage. We can therefore prepare any two-photon polarization state, ranging from product to maximally entangled states, as a superposition of two of the four Bell states.

All the presented results have been obtained with near perfect entanglement qualities, \textit{i.e.}, the Bell inequalities have been violated for all chosen bandwidths by more than 40 standard deviations, without noise subtraction. These results have also been made possible by the high brightness of our source, which is mainly due to a high-efficiency waveguide photon-pair generator, and optimized losses in the presented configuration. We have also outlined suggestions for future improvements that could lead to a potential brightness increase of almost one order of magnitude. These modifications are currently being implemented on our setup. 

We believe that our scheme is a good candidate for a broad range of quantum experiments, ranging from fundamental quantum optics to quantum network applications. In the latter framework, our photon spectral bandwidths of 540\,MHz and 25\,MHz are already suitable for several quantum memory strategies based on hot atomic vapors and solid state devices. In order to adapt the wavelengths of the entangled photons to the absorption resonances of the corresponding memory devices, non-linear wavelength conversion processes such as sum frequency generation and difference-frequency generation have been proven to be very efficient strategies~\cite{tanzilli_2005,curtz_2010}.

\section{Acknowledgements}
The authors would like to thank ID Quantique, Scontel, AOS GmbH, Draka Comteq, and Opton Laser International (Toptica Photonics) for technical support. The authors acknowledge financial support from the CNRS, the University Nice Sophia Antipolis, l'Agence Nationale de la Recherche for the 'e-QUANET' project (grant agreement ANR-09-BLAN-0333-01), the European ICT-2009.8.0 FET open program for the 'QUANTIP' project (grant agreement 244026), le Minist\`ere de l'Enseignement Sup\'erieur et de la Recherche (MESR), la Direction G\'en\'erale de l'Armement (DGA), the Majlis Amanah Rakyat (MARA), and le Conseil R\'egional PACA.


\begin{thebibliography}{10}

\bibitem{scarani_2009}
V.~Scarani, H.~{Bechmann-Pasquinucci}, N.~J. Cerf, M.~Duscaronek,
  N.~L\"utkenhaus, and M.~Peev.
\newblock The security of practical quantum key distribution.
\newblock {\em Rev. Mod. Phys.}, 81(3):1301, 2009.

\bibitem{Yoshino_HSQKD_2012}
K.-I. Yoshino, M.~Fujiwara, A.~Tanaka, S.~Takahashi, Y.~Nambu, A.~Tomita,
  S.~Miki, T.~Yamashita, Z.~Wang, M.~Sasaki, and A.~Tajima.
\newblock High-speed wavelength-division multiplexing quantum key distribution
  system.
\newblock {\em Opt. Lett.}, 37:223--225, 2012.

\bibitem{aboussouan_QR_2010}
P.~Aboussouan, O.~Alibart, D.~B. Ostrowsky, P.~Baldi, and S.~Tanzilli.
\newblock High-visibility two-photon interference at a telecom wavelength using
  picosecond-regime separated sources.
\newblock {\em Phys. Rev. A}, 81(2):021801, 2010.

\bibitem{McMillan_2photSeparated_2013}
A.~R. McMillan, L.~LabontÈ, B.~Bell, A.~Clark, A.~Martin, O.~Alibart,
  W.~Wadsworth, J.~R. Rarity, and S.~Tanzilli.
\newblock Two-photon interference between disparate sources for quantum
  networking.
\newblock {\em Sci. Rep.}, 3:2032, 2013.

\bibitem{lvovsky_OQM_2009}
A.~I. Lvovsky, B.~C. Sanders, and W.~Tittel.
\newblock Optical quantum memory.
\newblock {\em Nat. Photonics}, 3(12):706--714, 2009.

\bibitem{Sangouard_Rep_2011}
N.~Sangouard, C.~Simon, H.~de~Riedmatten, and N.~Gisin.
\newblock Quantum repeaters based on atomic ensembles and linear optics.
\newblock {\em Rev. Mod. Phys.}, 83:33--80, 2011.

\bibitem{aspect_experimental_1982}
A.~Aspect, P.~Grangier, and G.~Roger.
\newblock Experimental realization of {Einstein-Podolsky-Rosen-Bohm}
  gedankenexperiment: A new violation of {B}ell's inequalities.
\newblock {\em Phys. Rev. Lett.}, 49:91--94, 1982.

\bibitem{Tittel_2001}
W.~Tittel and G.~Weihs.
\newblock Photonic entanglement for fundamental tests and quantum
  communication.
\newblock {\em Quant. Inf. Comp.}, 1:3--56, 2001.

\bibitem{Tittel_Bell10_1998}
W.~Tittel, J.~Brendel, H.~Zbinden, and N.~Gisin.
\newblock Violation of {B}ell inequalities by photons more than 10 km apart.
\newblock {\em Phys. Rev. Lett.}, 81:3563--3566, 1998.

\bibitem{Weihs_Bell_1998}
G.~Weihs, T.~Jennewein, C.~Simon, H.~Weinfurter, and A.~Zeilinger.
\newblock Violation of bell's inequality under strict einstein locality
  conditions.
\newblock {\em Phys. Rev. Lett.}, 81:5039--5043, 1998.

\bibitem{Ma_2012}
X.~Ma, S.~Zotter, J.~Kofler, R.~Ursin, T.~Jennewein, C.~Brukner, and
  A.~Zeilinger.
\newblock Experimental delayed-choice entanglement swapping.
\newblock {\em Nat. Phys.}, 8:480--485, 2012.

\bibitem{peruzzo_2012}
A.~Peruzzo, P.~J. Shadbolt, N.~Brunner, S.~Popescu, and J.~L. O'Brien.
\newblock A quantum delayed choice experiment.
\newblock {\em Science}, 338:634, 2012.

\bibitem{kaiser_2012}
F.~Kaiser, T.~Coudreau, P.~Milman, D.~B. Ostrowsky, and S.~Tanzilli.
\newblock Entanglement-enabled delayed-choice experiment.
\newblock {\em Science}, 338:637, 2012.

\bibitem{Lvovsky_MicroMacro_2013}
A.~I. Lvovsky, R.~Ghobadi, A.~Chandra, A.~S. Prasad, and C.~Simon.
\newblock {Observation of micro?macro entanglement of light}.
\newblock {\em Nat. Phys.}, 9(8):541--544, 2013.

\bibitem{Bruno_MicroMacro_2013}
N.~Bruno, a.~Martin, P.~Sekatski, N.~Sangouard, R.~T. Thew, and N.~Gisin.
\newblock {Displacement of entanglement back and forth between the micro and
  macro domains}.
\newblock {\em Nat. Phys.}, 9(8):545--548, 2013.

\bibitem{lloyd_long_2001}
S.~Lloyd, M.~S. Shahriar, J.~H. Shapiro, and P.~R. Hemmer.
\newblock Long distance, unconditional teleportation of atomic states via
  complete {B}ell state measurements.
\newblock {\em Phys. Rev. Lett.}, 87:167903, 2001.

\bibitem{landry_quantum_2007}
O~Landry, J.~A.~W. van Houwelingen, A.~Beveratos, H.~Zbinden, and N.~Gisin.
\newblock Quantum teleportation over the {S}wisscom telecommunication network.
\newblock {\em J. Opt. Soc. Am. B}, 24:398, 2007.

\bibitem{Ursin_entanglement_2007}
R.~Ursin, F.~Tiefenbacher, T.~Schmitt-Manderbach, H.~Weier, T.~Scheidl,
  M.~Lindenthal, B.~Blauensteiner, T.~Jennewein, J.~Perdigues, P.~Trojek,
  B.~\"{O}mer, M.~F\"{u}rst, M.~Meyenburg, J.~G. Rarity, Z.~Sodnik,
  C.~Barbieri, H.~Weinfurter, and A.~Zeilinger.
\newblock Entanglement-based quantum communication over 144 km.
\newblock {\em Nat. Phys.}, 3:481--486, 2007.

\bibitem{Hubel_2007}
H.~H\"{u}bel, M.~R. Vanner, T.~Lederer, B.~Blauensteiner, T.~Lor\"{u}nser,
  A.~Poppe, and A.~Zeilinger.
\newblock High-fidelity transmission of polarization encoded qubits from an
  entangled source over 100 km of fiber.
\newblock {\em Opt. Express}, 15(12):7853, 2007.

\bibitem{Fasel2004a}
S.~Fasel, N.~Gisin, G.~Ribordy, and H.~Zbinden.
\newblock Quantum key distribution over 30 km of standard fiber using
  energy-time entangled photon pairs: a comparison of two chromatic dispersion
  reduction methods.
\newblock {\em Eur. Phys. J. D}, 30:143--148, 2004.

\bibitem{Tanzilli_Genesis_2012}
S.~Tanzilli, A.~Martin, F.~Kaiser, M.~P. De~Micheli, O.~Alibart, and D.~B.
  Ostrowsky.
\newblock On the genesis and evolution of quantum integrated optics.
\newblock {\em Laser \& Photon. Rev.}, 6:115--143, 2012.

\bibitem{kaiserI_2012a}
F.~Kaiser, A.~Issautier, L.~A. Ngah, O.~Danila, H.~Herrmann, W.~Sohler,
  A.~Martin, and S.~Tanzilli.
\newblock High-quality polarization entanglement state preparation and
  manipulation in standard telecommunication channels.
\newblock {\em New J. Phys.}, 14:085015, 2012.

\bibitem{Herrmann2013a}
H.~Herrmann, X.~Yang, A.~Thomas, A.~Poppe, W.~Sohler, and C.~Silberhorn.
\newblock {Post-selection free, integrated optical source of non-degenerate,
  polarization entangled photon pairs}.
\newblock {\em Opt. Express}, 21:27981, 2013.

\bibitem{Kaiser_type0_2013}
F.~Kaiser, A.~Issautier, L.~A. Ngah, O.~Alibart, A.~Martin, and S.~Tanzilli.
\newblock A versatile source of polarization entanglement for quantum network
  applications.
\newblock {\em Laser Phys. Lett.}, 10:045202, 2013.

\bibitem{Fulconis_PolarFiber_2007}
J.~Fulconis, O.~Alibart, J.~L. O'Brien, W.~J. Wadsworth, and J.~G. Rarity.
\newblock Nonclassical interference and entanglement generation using a
  photonic crystal fiber pair photon source.
\newblock {\em Phys. Rev. Lett.}, 99:120501, 2007.

\bibitem{Fang_PolarFiber_2013}
B.~Fang, O.~Cohen, and V.~O. Lorenz.
\newblock Polarization-entangled photon-pair generation in commercial-grade
  polarization-maintaining fiber.
\newblock {\em J. Opt. Soc. Am. B}, 31:277--281, 2014.

\bibitem{Fiorentino2008}
M.~Fiorentino and R.~G. Beausoleil.
\newblock Compact sources of polarization-entangled photons.
\newblock {\em Opt. Express}, 16:20149, 2008.

\bibitem{piro_2009}
N.~Piro, A.~Haase, M.~W. Mitchell, and J.~Eschner.
\newblock An entangled photon source for resonant single-photon-single-atom
  interaction.
\newblock {\em J. Phys. B: At. Mol. Opt. Phys.}, 42:114002, 2009.

\bibitem{Wang2004}
H.~Wang, T.~Horikiri, and T.~Kobayashi.
\newblock {Polarization-entangled mode-locked photons from cavity-enhanced
  spontaneous parametric down-conversion}.
\newblock {\em Phys. Rev. A}, 70:043804, 2004.

\bibitem{Kuklewicz_2006}
C.~E.~Kuklewicz, F.~N.~C.~Wong, and J.~H.~Shapiro.
\newblock {Time-Bin-Modulated Biphotons from Cavity-Enhanced Down-Conversion}.
\newblock {\em Phys. Rev. Lett.}, 97:223601, 2006.

\bibitem{bao_2008}
{X.-H.} Bao, Y.~Qian, J.~Yang, H.~Zhang, {Z.-B.} Chen, T.~Yang, and {J.-W.}
  Pan.
\newblock Generation of {Narrow-Band} {Polarization-Entangled} photon pairs for
  atomic quantum memories.
\newblock {\em Phys. Rev. Lett.}, 101(19):190501, 2008.

\bibitem{dousse_2010}
A.~Dousse, J.~Suffczynski, A.~Beveratos, O.~Krebs, A.~Lemaitre, I.~Sagnes,
  J.~Bloch, P.~Voisin, and P.~Senellart.
\newblock Ultrabright source of entangled photon pairs.
\newblock {\em Nature}, 466(7303):217--220, 2010.

\bibitem{dudin_2010}
Y.~O. Dudin, A.~G. Radnaev, R.~Zhao, J.~Z. Blumoff, T.~A.~B. Kennedy, and
  A.~Kuzmich.
\newblock Entanglement of light-shift compensated atomic spin waves with
  telecom light.
\newblock {\em Phys. Rev. Lett.}, 105:260502, 2010.

\bibitem{yan_2011}
H.~Yan, S.~Zhang, J.~F. Chen, M.~M.~T. Loy, G.~K.~L. Wong, and S.~Du.
\newblock Generation of narrow-band hyperentangled nondegenerate paired
  photons.
\newblock {\em Phys. Rev. Lett.}, 106:033601, 2011.

\bibitem{Martin_TB_2013}
A.~Martin, F.~Kaiser, Vernier A., A.~Beveratos, V.~Scarani, and S.~Tanzilli.
\newblock Cross time-bin photonic entanglement for quantum key distribution.
\newblock {\em Phys. Rev. A}, 87:020301(R), 2013.

\bibitem{xavier_2009}
G.~B. Xavier, N.~Walenta, G.~Vilela~de Faria, G.~P. Tempor\~{a}o, N.~Gisin,
  H.~Zbinden, and J.~P. von~der Weid.
\newblock Experimental polarization encoded quantum key distribution over
  optical fibres with real-time continuous birefringence compensation.
\newblock {\em New J. Phys.}, 11(4):045015, 2009.

\bibitem{reim_2010}
K.~F. Reim, J.~Nunn, V.~O. Lorenz, B.~J. Sussman, K.~C. Lee, N.~K. Langford,
  D.~Jaksch, and I.~A. Walmsley.
\newblock Towards high-speed optical quantum memories.
\newblock {\em Nat. Photonics}, 4(4):218--221, 2010.

\bibitem{clausen_2011}
C.~Clausen, I.~Usmani, F.~Bussieres, N.~Sangouard, M.~Afzelius,
  H.~de~Riedmatten, and N.~Gisin.
\newblock Quantum storage of photonic entanglement in a crystal.
\newblock {\em Nature}, 469(7331):508--511, 2011.

\bibitem{saglamyurek_2011}
E.~Saglamyurek, N.~Sinclair, J.~Jin, J.~A. Slater, D.~Oblak, F.~Bussieres,
  M.~George, R.~Ricken, W.~Sohler, and W.~Tittel.
\newblock Broadband waveguide quantum memory for entangled photons.
\newblock {\em Nature}, 469(7331):512--515, 2011.

\bibitem{tanji_2009}
H.~Tanji, S.~Ghosh, J.~Simon, B.~Bloom, and V.~Vuleti\ifmmode~\acute{c}\else
  \'{c}\fi{}.
\newblock Heralded single-magnon quantum memory for photon polarization states.
\newblock {\em Phys. Rev. Lett.}, 103:043601, 2009.

\bibitem{piro_2011}
N.~Piro, F.~Rohde, C.~Schuck, M.~Almendros, J.~Huwer, J.~Ghosh, A.~Haase,
  M.~Hennrich, F.~Dubin, and J.~Eschner.
\newblock Heralded single-photon absorption by a single atom.
\newblock {\em Nat. Phys.}, 7:17--20, 2011.

\bibitem{Franson_Bell_1989}
J.~D Franson.
\newblock Bell inequality for position and time.
\newblock {\em Phys. Rev. Lett.}, 62:2205--2208, 1989.

\bibitem{chanvillard_SPE_2000}
L.~Chanvillard, P.~Aschi\'eri, P.~Baldi, D.~B. Ostrowsky, M.~P. De~Micheli,
  L.~Huang, and D.~J. Bamford.
\newblock Soft proton exchange on periodically poled {L}i{N}b{O}$_3$: A simple
  waveguide fabrication process for highly efficient nonlinear interactions.
\newblock {\em Appl. Phys. Lett.}, 76(9):1089--1091, 2000.

\bibitem{Tanzilli_PPLNW_2002}
S.~Tanzilli, W.~Tittel, H.~De~Riedmatten, H.~Zbinden, P.~Baldi, M.~P.
  De~Micheli, D.~B. Ostrowsky, and N.~Gisin.
\newblock {PPLN} waveguides for quantum communication.
\newblock {\em Eur. J. Phys. D}, 18:155--160, 2002.

\bibitem{Halder_HighCoh_2008}
M.~Halder, A.~Beveratos, R.~T. Thew, C.~Jorel, H.~Zbinden, and N.~Gisin.
\newblock High coherence photon pair source for quantum communication.
\newblock {\em New J. Phys.}, 10, 2008.

\bibitem{martin_2012}
A.~Martin, J.-L. Smirr, F.~Kaiser, E.~Diamanti, A.~Issautier, O.~Alibart,
  R.~Frey, I.~Zaquine, and S.~Tanzilli.
\newblock Analysis of elliptically polarized maximally entangled states for
  {B}ell inequality tests.
\newblock {\em Laser Phys.}, 22:1105--1112, 2012.

\bibitem{leviton_2008}
D.~B. Leviton and B.~J. Frey.
\newblock Temperature-dependent absolute refractive index measurements of
  synthetic fused silica.
\newblock {\em arXiv:0805.0091}, 2008.

\bibitem{Altepeter_experimental_2005}
J.~B. Altepeter, E.~R. Jeffrey, P.~G. Kwiat, S.~Tanzilli, N.~Gisin, and
  A.~Ac\'in.
\newblock Experimental methods for detecting entanglement.
\newblock {\em Phys. Rev. Lett.}, 95:033601, 2005.

\bibitem{bell_1964}
{J. S.} Bell.
\newblock On the {Einstein-Podolsky-Rosen} paradox.
\newblock {\em Physics}, 1:195--200, 1964.

\bibitem{clauser_1969}
J.~F. Clauser, M.~A. Horne, A.~Shimony, and R.~A. Holt.
\newblock Proposed experiment to test local {Hidden-Variable} theories.
\newblock {\em Phys. Rev. Lett.}, 23(15):880--884, 1969.

\bibitem{Tittel_BellGeneral_1999}
W.~Tittel, J.~Brendel, H.~Zbinden, and N.~Gisin.
\newblock Long distance {B}ell-type tests using energy-time entangled photons.
\newblock {\em Phys. Rev. A}, 59:4150--4163, 1999.

\bibitem{Tapster1998}
P.~R. Tapster and J.~G. Rarity.
\newblock {Photon statistics of pulsed parametric light}.
\newblock {\em J. Mod. Opt.}, 45:595--604, 1998.

\bibitem{Monteiro_Pure_2014}
F.~Monteiro, A.~Martin, B.~Sanguinetti, H.~Zbinden, and R.~T. Thew.
\newblock Narrowband photon pair source for quantum networks.
\newblock {\em To appear in Opt. Express}, 2014.
\newblock Eprint arXiv:13123832.

\bibitem{Scarani_2pairs_04}
H.~de~Riedmatten, V.~Scarani, I.~Marcikic, A.~Acin, W.~Tittel, H.~Zbinden, and
  N.~Gisin.
\newblock Two independent photon pairs versus four-photon parametric down
  conversion entangled states in parametric down conversion.
\newblock {\em J. Mod. Opt.}, 51:1637, 2004.

\bibitem{Sekatski2012}
P.~Sekatski, N.~Sangouard, F.~Bussi\`{e}res, C.~Clausen, N.~Gisin, and
  H.~Zbinden.
\newblock {Detector imperfections in photon-pair source characterization}.
\newblock {\em J. Phys. B}, 45:124016, 2012.

\bibitem{hadfield_2009}
R.~H. Hadfield.
\newblock Single-photon detectors for optical quantum information applications.
\newblock {\em Nat. Photonics}, 3(12):696--705, 2009.

\bibitem{castaldini_2009}
D.~Castaldini, P.~Bassi, P.~Aschieri, S.~Tascu, M.~De Micheli, and P.~Baldi.
\newblock High performance mode adapters based on segmented {SPE:LiNbO}$_3$
  waveguides.
\newblock {\em Opt. Express}, 17(20):17868--17873, 2009.

\bibitem{tanzilli_2005}
S.~Tanzilli, W.~Tittel, M.~Halder, O.~Alibart, P.~Baldi, N.~Gisin, and
  H.~Zbinden.
\newblock A photonic quantum information interface.
\newblock {\em Nature}, 437(7055):116--120, 2005.

\bibitem{curtz_2010}
N.~Curtz, R.~Thew, C.~Simon, N.~Gisin, and H.~Zbinden.
\newblock Coherent frequency-down-conversion interface for quantum repeaters.
\newblock {\em Opt. Express}, 18:22099--22104, 2010.

\end{thebibliography}
\end{document}